\documentclass[conference]{IEEEtran}
\usepackage{amsmath,amssymb,amsthm}
\usepackage{graphicx}
\usepackage{psfrag}

\usepackage{mathptmx}
\DeclareMathAlphabet{\mathcal}{OMS}{cmsy}{m}{n}

\allowdisplaybreaks[2]

\title{Connection between Annealed Free Energy and Belief Propagation on Random Factor Graph Ensembles}

\author{
\IEEEauthorblockN{Ryuhei Mori}
\IEEEauthorblockA{
Graduate School of Informatics,
Kyoto University\\
Kyoto, 606--8501, Japan\\
Email: rmori@sys.i.kyoto-u.ac.jp
}
}

\theoremstyle{plain}
\newtheorem{theorem}{Theorem}
\newtheorem{lemma}[theorem]{Lemma}
\newtheorem{corollary}[theorem]{Corollary}
\theoremstyle{definition}

\newtheorem{remark}[theorem]{Remark}

\begin{document}
\maketitle
\begin{abstract}
Recently, Vontobel showed the relationship between Bethe free energy and annealed free energy for protograph factor graph ensembles.
In this paper, annealed free energy of any random regular, irregular and Poisson factor graph ensembles are connected to Bethe free energy.
The annealed free energy is expressed as the solution of maximization problem whose stationary condition equations coincide with equations of belief propagation
since the contribution to partition function of particular type of variable and factor nodes has similar form of minus Bethe free energy.
It gives simple derivation of replica symmetric solution.
As consequence, it is shown that on replica symmetric ansatz, replica symmetric solution and annealed free energy are equal for regular ensemble.
\end{abstract}

\section{Introduction}
In the context of statistical physics, free energy of disordered system is central interest.
In information theory, the a posteriori distribution of low-density parity-check (LDPC) codes can be regarded as
Boltzmann-Gibbs distributions on sparse factor graphs
whose free energy is related to the conditional entropy of codewords under a received vector~\cite{macris2007griffith}.
In computer science, constraint satisfaction problems (CSPs) which can be expressed by sparse factor graphs are important theoretical objects.
Relation between phase transition phenomenon and free energy of randomized CSPs has also been considered well~\cite{van2002random}, \cite{mezard2009information}. 

In this paper, we deal with calculation of \textit{annealed free energy} of random sparse factor graph ensemble on finite alphabet.
Although in many cases \textit{quenched free energy} gives meaningful result
e.g., conditional entropy of LDPC codes~\cite{macris2007griffith}, phase transition point of random CSPs~\cite{van2002random},
the calculation of quenched free energy is often difficult without \textit{replica method} which is mathematically nonrigorous but powerful tool of statistical physics.
Annealed free energy is also important quantity since it can be used for bound of quenched free energy and is required in the replica method.

For many cases~\cite{di2005weight}, in the calculation of annealed and quenched free energy, fixed point equations of
belief propagation (BP) and its density evolution (DE) appear, respectively.
However, the relationship between BP (DE) and annealed (quenched) free energy has not been well understood.
Recently, Vontobel show the relationship between Bethe free energy of protograph ensemble and its annealed free energy~\cite{vontobel2010counting}.
From this result, we can connect BP and annealed free energy
since BP equation is equivalent to stationary condition of Bethe free energy~\cite{yedidia2005constructing}.

The main result of this paper is derivation of annealed free energy of any random regular, irregular and Poisson factor graph ensembles by using BP equations.
The derivation of annealed free energy gives the simple derivation of replica symmetric solution.
It is shown that if the replica symmetric assumption is correct, annealed and quenched free energy are equal for any regular ensembles. 


\section{Factor graph, Gibbs free energy and Bethe approximation}
In this paper, we deal with factor graph which is bipartite graph representing probability distribution~\cite{yedidia2005constructing}, \cite{mezard2009information}.
Let us consider bipartite graph consists of $N$ variable nodes and $M$ factor nodes.
Let $\mathcal{X}$ be alphabet which is common domain of variables.
For each factor node $a$, there is a function $f_a: \mathcal{X}^{r_a}\to\mathbb{R}_{\ge 0}$ where $r_a$ denotes the degree of $a$.
The factor graph represents the following distribution $p$ on $\mathcal{X}^N$.
\begin{equation*}
p(\pmb{x}) = \frac1Z \prod_a f_a(\pmb{x}_{\partial a})
\end{equation*}
where
\begin{equation*}
Z := \sum_{\pmb{x}} \prod_a f_a(\pmb{x}_{\partial a})
\end{equation*}
is constant for normalization, i.e., $\sum_{\pmb{x}} p(\pmb{x}) = 1$.
Here, $\pmb{x}_{\partial a}$ denotes value of variable nodes connecting a factor node $a$.
In the context of statistical mechanics, $Z$ is called partition function and $-\log Z$ is called Helmholtz free energy.

When $N$ is large, calculation of $Z$ requires large computational complexity.
Hence, the approximation of $p$ by simple distribution $q$ is often introduced.
The following method of approximation is written in~\cite{yedidia2005constructing}.
For the criteria of approximation, Kullback-Leibler divergence is used.
\begin{align*}
D(q\| p) &:= \sum_{\pmb{x}}q(\pmb{x})\log\frac{q(\pmb{x})}{p(\pmb{x})}\\
&= \log Z - \sum_{\pmb{x}}\sum_a q(\pmb{x})\log f_a(\pmb{x}_{\partial a}) + \sum_{\pmb{x}} q(\pmb{x})\log q(\pmb{x})\\
&=: \log Z + \mathcal{U}(q) - \mathcal{H}(q) =: \log Z + \mathcal{F}_\text{Gibbs}(q)
\end{align*}
The quantity $\mathcal{U}(q)$, $\mathcal{H}(q)$ and $\mathcal{F}_\text{Gibbs}(q)$ are called internal energy, entropy and Gibbs free energy, respectively.

The approximation using $q(\pmb{x})$ which is factorized as $\prod_{i=1}^Nq_i(x_i)$, i.e., $x_i$ are independent,
is called mean field approximation.
The approximation using $q(\pmb{x})$ which is represented as
\begin{equation*}
q(\pmb{x}) = \frac{\prod_a b_a(\pmb{x}_{\partial a})}{\prod_i b_i(x_i)^{l_i-1}}
\end{equation*}
is called \textit{Bethe approximation} where $i$ and $a$ represent indices of variable nodes and factor nodes, respectively,
and where $l_i$ denotes degree of variable node $i$.
For Bethe approximation, 
Bethe average energy and Bethe entropy are defined as
\begin{align}
\mathcal{U}_\text{Bethe}(b_a) &:= -\sum_{a}\sum_{\pmb{x}_{\partial a}} b_a(\pmb{x}_{\partial a}) \log f_a(\pmb{x}_{\partial a})\nonumber\\
\mathcal{H}_\text{Bethe}(b_i,b_a) &:= -\sum_{a}\sum_{\pmb{x}_{\partial a}} b_a(\pmb{x}_{\partial a}) \log b_a(\pmb{x}_{\partial a})\nonumber\\
&\quad +\sum_{i}\sum_{x_i} (l_i-1)b_i(x_i) \log b_i(x_i)\label{eq:Betheent}
\end{align}
respectively.
Bethe free energy is defined as $\mathcal{F}_\text{Bethe}(b_i,b_a):=\mathcal{U}_\text{Bethe}(b_a)-\mathcal{H}_\text{Bethe}(b_i,b_a)$.
In order to obtain good Bethe approximation, minimization of Bethe free energy is considered since Bethe free energy is analogy of
Gibbs free energy, whose minimization is equivalent to minimization of the Kullback-Leibler divergence.
When we assume constraints, $\sum_{i}b_i(x_i)=1$ for all variable nodes $i$, $\sum_{\pmb{x}_{\partial a}}b_a(\pmb{x}_{\partial a}) = 1$ for all factor nodes $a$,
and $\sum_{\pmb{x}_{\partial a}, x_i=x}b_a(\pmb{x}_{\partial a})=b_i(x)$ for all factor nodes $a$ and variable nodes $i\in\partial a$,
the stationary condition of Lagrangian is equivalent to condition of fixed point of BP~\cite{yedidia2005constructing}.



\section{Annealed free energy of random regular factor graph ensembles}
In this paper, we mainly deal with random regular factor graph ensembles.
Results for regular ensembles can be generalized straightforwardly to irregular and Poisson ensembles.
Let $l$ and $r$ be degrees of variable and factor nodes of regular factor graph ensembles, respectively.
Let $\mathbb{E}[\cdot]$ denote the expectation on random connection of edges.
Two quantities $\mathbb{E}[\log Z]$ and $\log\mathbb{E}[Z]$ are called quenched and annealed free energy, respectively.
The main purpose of this paper is calculation of $\lim_{N\to\infty}1/N\log\mathbb{E}[Z]$ where $N$ denotes the number of variable nodes.
The essential idea of calculation is type classification of the contribution to partition function~\cite{vontobel2010counting}.
Let \textit{variable-type} $v$ denote the type of variable nodes, i.e., there exists $v(x)$ variable nodes of value $x\in\mathcal{X}$.
Let \textit{factor-type} $u$ denote the type of factor nodes, in which the value of factor nodes is regarded as the values of variable nodes connects to the factor nodes,
i.e., there exists $u(\pmb{x})$ factor nodes connecting variable nodes of value $\pmb{x}\in\mathcal{X}^r$.
In this paper, for simplicity, factors $f_a(\pmb{x}_{\partial a})$ do not depend on factor node $a$, and written as $f(\pmb{x}_{\partial a})$.
Let $Z(v,u)$ be the contribution of assignments with variable-type $v$ and factor-type $u$,
and $N(v,u)$ be the number of assignments with variable-type $v$ and factor-type $u$.
\begin{equation*}
Z = \sum_{v,u}Z(v,u)
=\sum_{v,u} N(v,u)\prod_{\pmb{x}\in\mathcal{X}^r}f(\pmb{x})^{u({\pmb{x}})}
\end{equation*}
In the sum, the types $v$ and $u$ have to satisfy the consistency condition
\begin{equation*}
\sum_{i=1}^r\sum_{\substack{\pmb{x}\setminus x_i\\x_i=z}} u(\pmb{x}) = lv(z).
\end{equation*}
The number $N(v,u)$ of assignments with variable-type $v$ and factor-type $u$ is
\begin{equation*}
\mathbb{E}[N(v,u)]
=
\binom{N}{\{v(x)\}_{x\in\mathcal{X}}}
\binom{\frac{l}{r}N}{\{u(\pmb{x})\}_{\pmb{x}\in\mathcal{X}^r}}
\frac{\prod_{x\in\mathcal{X}}
(v(x)l)!}{(Nl)!}.
\end{equation*}
Now, we consider the exponent of the contribution of types $\nu$ and $\mu$ where $\nu(x) := v(x)/N$ and $\mu(x) := u(x)/((l/r) N)$, respectively.
It holds
\begin{multline*}
\lim_{N\to\infty} \frac1N \log \mathbb{E}[Z(\nu,\mu)]\\
= \frac{l}{r}\mathcal{H}(\mu) - (l-1)\mathcal{H}(\nu)
+\frac{l}{r}\sum_{\pmb{x}\in\mathcal{X}^r} \mu(\pmb{x})\log f(\pmb{x})\\
=: -F_\text{Bethe}(\nu,\mu).
\end{multline*}
Hence,
\begin{equation*}
\lim_{N\to\infty} \frac1N \log \mathbb{E}[Z] = \max_{\nu,\mu} \left\{-F_\text{Bethe}(\nu,\mu)\right\}
\end{equation*}
where, $\nu$ and $\mu$ have to satisfy the following conditions.
\begin{align*}
\nu(x) &\ge 0, \forall x\in\mathcal{X},&
\mu(\pmb{x}) &\ge 0, \forall \pmb{x}\in\mathcal{X}^r\\
\sum_{x\in\mathcal{X}} \nu(x) &= 1,&
\sum_{\pmb{x}\in\mathcal{X}^r} \mu(\pmb{x}) &= 1,
\end{align*}
\begin{equation*}
\frac1{r}\sum_{i=1}^r\sum_{\substack{\pmb{x}\setminus x_i\\ x_i=z}}\mu(\pmb{x}) = \nu(z),\forall z\in\mathcal{X}.
\end{equation*}
The last condition is for the consistency between $\nu$ and $\mu$.
The above maximization problem is similar to the minimization problem of Bethe free energy.
Hence, we can easily understand that the stationary condition is similar to the fixed point equation of BP.
The Lagrangian of the maximization problem is
\begin{multline}\label{eq:Lag}
L(\nu,\mu;\lambda,\rho, \tau) = -F_\text{Bethe}(\nu,\mu)\\
+ \lambda\left(\sum_{x\in\mathcal{X}}\nu(x)-1\right)
+ \frac{l}{r}\rho\left(\sum_{\pmb{x}\in\mathcal{X}^r}\mu(\pmb{x})-1\right)\\
+\sum_{z\in\mathcal{X}}\tau(z)\left(\frac{l}{r}\sum_{i=1}^r\sum_{\substack{\pmb{x}\setminus x_i\\ x_i=z}}\mu(\pmb{x}) - l\nu(z)\right).
\end{multline}
\begin{lemma}\label{lem:stat}
The stationary condition of~\eqref{eq:Lag} is
\begin{align*}
\nu(x) &\propto m_{f\to v}(x)^l\\
\mu(\pmb{x}) &\propto f(\pmb{x})\prod_{i=1}^r m_{v\to f}(x_i)
\end{align*}
where
\begin{align}
m_{v\to f}(x)&\propto m_{f\to v}(x)^{l-1}\label{eq:saddles}\\
m_{f\to v}(x)&\propto \sum_{i=1}^r\sum_{\substack{\pmb{x}\setminus x_i\\x_i=x}}f(\pmb{x})\prod_{j=1, j\ne i}^r m_{v\to f}(x_j)\label{eq:saddlee}.
\end{align}
Here $m_{v\to f}(x)$ and $m_{f\to v}(x)$ are auxiliary functions satisfying
$\sum_{x\in\mathcal{X}} m_{v\to f}(x) = \sum_{x\in\mathcal{X}} m_{f\to v}(x) = 1$.
\end{lemma}
Proof is in Appendix~\ref{apdx:stat}.
If $f(\pmb{x})$ is invariant under permutation of $\pmb{x}$,
\eqref{eq:saddlee} is simply written as
\begin{equation*}
m_{f\to v}(x)\propto \sum_{\substack{\pmb{x}\setminus x_1\\x_1=x}}f(\pmb{x})\prod_{j=2}^r m_{v\to f}(x_j).
\end{equation*}

\vspace{-1em}
\begin{theorem}\label{thm:ann}
\begin{multline*}
\lim_{N\to\infty}\frac1N\log \mathbb{E}[Z]\\
 = \max_{(m_{v\to f}(x),m_{f\to v}(x))\in\mathcal{S}}\left\{\frac{l}{r}\log Z_f + \log Z_v - l\log Z_{fv}\right\}.
\end{multline*}
where $\mathcal{S}$ denotes the set of saddle points of the function in $\max$, and where
\begin{align*}
Z_{v} &:= \sum_{x} m_{f\to v}(x)^l\\
Z_{f} &:= \sum_{\pmb{x}} f(\pmb{x}) \prod_{i=1}^r m_{v\to f}(x_i)\\
Z_{fv} &:= \sum_{x} m_{f\to v}(x)m_{v\to f}(x).
\end{align*}
The conditions of saddle point are~\eqref{eq:saddles} and \eqref{eq:saddlee}.
\end{theorem}
Proof is in Appendix~\ref{apdx:ann}.

\begin{remark}
Assume that $\sum_{i=1}^r\sum_{\substack{\pmb{x}\setminus x_i\\x_i=x}} f(\pmb{x})$ is constant among all $x\in\mathcal{X}$.
Then, the uniform distributions $m_{v\to f}(x)$ and $m_{f\to v}(x)$ are a trivial fixed point.
Let $N_f:=\sum_{\pmb{x}}f(\pmb{x})$.
The contribution $Z(\nu,\mu)$ evaluated at uniform $\nu$ and $\mu$ is
\begin{equation}\label{eq:dr}
\frac{l}{r}\log\frac{N_f}{q^r} + \log \frac1{q^{l-1}} - l\log\frac1q
= \log q + \frac{l}{r}\log\frac{N_f}{q^r}.
\end{equation}
When $f(\pmb{x})\in\{0,1\}$, i.e., the problem is the CSP,
$Z$ is the number of solution and $N_f$ is the cardinality of $\{\pmb{x}\in\mathcal{X}^r\mid f(\pmb{x})=1\}$.
In this case, we call the quantity~\eqref{eq:dr} \textit{design rate}.
If the uniform $\nu$ and $\mu$ maximize $Z(\nu,\mu)$,
the expected number $\mathbb{E}[Z]$ of solution is about
\begin{equation*}
q^N\left(\frac{N_f}{q^r}\right)^{\frac{l}{r}N}.
\end{equation*}
Roughly speaking, this implies that all constraints are independent.
This solution is called \textit{paramagnetic solution} in~\cite{van2002random},
in the context of replica symmetric solution.
\end{remark}
The generalization for irregular and Poisson ensemble is in Appendix~\ref{apdx:irgpoi}.

\section{Contribution to partition function of fixed variable type}\label{sec:fvt}
We now consider the contribution to partition function of regular factor graph ensemble with fixed variable type.
More precisely, we consider 
$Z(v) := \sum_{u} Z(v,u)$.
It holds
\begin{equation}
\lim_{N\to\infty}\frac1N \log \mathbb{E}[Z(\nu)] = \max_{\mu} \left\{-F_\text{Bethe}(\nu,\mu)\right\}.
\label{eq:Bethe2}
\end{equation}
The function $-F_\text{Bethe}(\nu,\mu)$ is a concave function with respect to $\mu$.
Since the equality constraints are linear, the problem is essentially a maximization problem of a concave function without constraints.
Hence, it can be solved numerically by the Newton method.
\begin{lemma}\label{lem:vtypestat}
The stationary condition of~\eqref{eq:Bethe2} is
\begin{align}
\mu(\pmb{x}) &\propto f(\pmb{x})\prod_{i=1}^r m_{v\to f}(x_i)\nonumber
\end{align}
where
\begin{align}
\nu(x) &\propto h(x)m_{f\to v}(x)^l\label{eq:v}\\
m_{v\to f}(x)&\propto h(x)m_{f\to v}(x)^{l-1}\label{eq:mvf}\\
m_{f\to v}(x)&\propto \sum_{i=1}^r\sum_{\substack{\pmb{x}\setminus x_i\\x_i=x}}f(\pmb{x})\prod_{j=1, j\ne i}^r m_{v\to f}(x_j).\label{eq:mfv}
\end{align}
Here $m_{v\to f}(x)$, $m_{f\to v}(x)$ and $h(x)$ are auxiliary functions satisfying
$\sum_{x\in\mathcal{X}} m_{v\to f}(x) = \sum_{x\in\mathcal{X}} m_{f\to v}(x) = 1$,
and $h(x)\ge 0$.
\end{lemma}
Proof is in Appendix~\ref{apdx:vtypestat}.
Since $h(x)$ is arbitrary auxiliary function, $m_{f\to v}(x)^l$ in~\eqref{eq:v} and $m_{f\to v}(x)^{l-1}$ in~\eqref{eq:mvf},
can be replaced by $m_{f\to v}(x)^k$ and $m_{f\to v}(x)^{k-1}$, respectively for any $k\ge 1$.
Here, we chose $k=l$ since we can obtain the following simple result. 
The stationary condition for magnetic field model in Appendix~\ref{apdx:mag} and Lemma~\ref{lem:vtypestat} are the similar
although while in the problem for magnetic field,
$h(x)$ is given and $\nu(x)$ is variable, in this problem, $h(x)$ is variable and $\nu(x)$ is given.
\begin{lemma}\label{lem:vtype}
\begin{align}
&\lim_{N\to\infty}\frac1N\log \mathbb{E}[Z(\nu)]\nonumber\\
&\; = \max_{(m_{f\to v}(x),m_{v\to f}(x),h(x))\in\mathcal{S}} \bigg\{\frac{l}{r}\log Z_f + \log Z_v - l\log Z_{fv}\nonumber\\
&\qquad - \sum_{x}\nu(x)\log h(x)\bigg\}\label{eq:vtype}\\
&\; = \max_{(m_{f\to v}(x),m_{v\to f}(x))\in\mathcal{S}}\left\{\frac{l}{r}\log Z_f + \sum_{x}\nu(x)\log Z_v(x) - l\log Z_{fv}\right\}\nonumber\\
&\qquad + \mathcal{H}(\nu).\nonumber
\end{align}
where $\mathcal{S}$ denotes the set of saddle points of the function in $\max$, and where
\begin{align*}
Z_{f} &:= \sum_{\pmb{x}} f(\pmb{x}) \prod_{i=1}^r m_{v\to f}(x_i)\\
Z_{v}(x) &:= m_{f\to v}(x)^l\\
Z_{v} &:= \sum_{x} h(x)Z_v(x)\\
Z_{fv} &:= \sum_{x} m_{f\to v}(x)m_{v\to f}(x).
\end{align*}
The conditions of saddle point are~\eqref{eq:v}, \eqref{eq:mvf} and \eqref{eq:mfv}.
\end{lemma}

The annealed free energy of magnetic field model in Appendix~\ref{apdx:mag} is obtained by the Legendre transform of the above result.
It can be easily verified from~\eqref{eq:vtype}.

While the maximization problem~\eqref{eq:Bethe2} can be solved by the Newton method, Lemma~\ref{lem:vtypestat} gives the efficient algorithm.
First, $\{m_{f\to v}^{(0)}(x)\}_{x\in\mathcal{X}}$ are initialized. Then, messages are updated by
\begin{align*}
m_{v\to f}^{(t+1)}(x) &\propto \frac{\nu(x)}{m_{f\to v}^{(t)}(x)}\\
m_{f\to v}^{(t)}(x)&\propto \sum_{i=1}^r\sum_{\substack{\pmb{x}\setminus x_i\\x_i=x}}f(\pmb{x})\prod_{j=1, j\ne i}^r m_{v\to f}^{(t)}(x_j)
\end{align*}
iteratively.
After sufficient iterations,
messages are substituted to
\begin{equation*}
l\left(\frac1{r}\log Z_f + \sum_{x}\nu(x)\log m_{f\to v}(x) - \log Z_{fv}\right)
+H(\nu).
\end{equation*}
Note that the degree $l$ of variable nodes does not appear in the iterations and only appear as the factor of the first term in the last equation.
This algorithm does not necessarily converges.
The example of problem for which the above BP-like algorithm does not converge is shown in Section~\ref{sec:app}.

\section{Moment of partition function and replica method}
In this section, we deal with moments of partition function which is useful for some purposes.
One of the most successful result of use of moment is the second moment method i.e.,
for nonnegative random variable $Z$,
$P(Z>0)\ge\mathbb{E}[Z]^2/\mathbb{E}[Z^2]$.
Using this method, lower bound of SAT-UNSAT threshold is obtained~\cite{achlioptas2003threshold}.
The other use of moment is the replica method which is not rigorous but powerful tool of statistical physics for calculation of quenched free energy.
The basic idea of the replica method is representation of $\mathbb{E}[\log Z]$ as the derivative $(\partial \log \mathbb{E}[Z^n])/\partial n|_{n=0}$.
It holds
\begin{equation*}
\lim_{N\to\infty}\frac1N\mathbb{E}[\log Z] 
=\lim_{N\to\infty}\frac1N\lim_{n\to 0}\frac{\log\mathbb{E}[Z^n]}{n}.
\end{equation*}
If the exchange of the limits is admissible,
\begin{equation}
\lim_{N\to\infty}\frac1N\mathbb{E}[\log Z] 
=\lim_{n\to 0}\frac1n\lim_{N\to\infty}\frac1N \log\mathbb{E}[Z^n].
\label{eq:replica}
\end{equation}
In the replica method,
\begin{equation}
\lim_{N\to\infty}\frac1N \log\mathbb{E}[Z^n]\label{eq:mom}
\end{equation}
have to be evaluated.
Usually, \eqref{eq:mom} is evaluated only for $n\in\mathbb{N}$ such that dependence on $n$ is analytic.
Then, the right-hand side of~\eqref{eq:replica} is evaluated by ignoring that $n$ should be natural number~\cite{mezard2009information}.

Since $Z^n$ can be regarded as partition function of factor graph on alphabet $\mathcal{X}^n$ and factor $\prod_{i=1}^n f(\pmb{x}^{(i)})$,
the exponent of moment is also calculated in the same way.
Here, $\pmb{x}^{(i)}\in\mathcal{X}^r$ denotes vector $(\pmb{x}_1^{(i)},\dotsc,\pmb{x}_r^{(i)})$
 where $\pmb{x}_j$ is $j$-th elements of $\pmb{x}\in(\mathcal{X}^n)^r$ and $\pmb{x}_j^{(i)}$ denotes $i$-th element of $\pmb{x}_j\in\mathcal{X}^n$.
\begin{corollary}
\begin{multline}\label{eq:moment}
\lim_{N\to\infty}\frac1N\log \mathbb{E}[Z^n]\\
 = \max_{(m_{f\to v}(\pmb{x}),m_{v\to f}(\pmb{x}))\in\mathcal{S}}\left\{\frac{l}{r}\log Z_f + \log Z_v - l\log Z_{fv}\right\}
\end{multline}
where $\mathcal{S}$ denotes the set of saddle points of the function in $\max$, and where
\begin{align*}
Z_{v} &:= \sum_{\pmb{x}\in\mathcal{X}^n} m_{f\to v}(\pmb{x})^l\\
Z_{f} &:= \sum_{\pmb{x}\in(\mathcal{X}^n)^r} \left(\prod_{j=1}^nf(\pmb{x}^{(j)})\right) \prod_{i=1}^r m_{v\to f}(\pmb{x}_i)\\
Z_{fv} &:= \sum_{\pmb{x}\in\mathcal{X}^n} m_{f\to v}(\pmb{x})m_{v\to f}(\pmb{x}).
\end{align*}
\end{corollary}
The essentially same result for LDPC codes was obtained in~\cite{montanari2001glassy} (Eq. (5.2)).
%
In~\cite{montanari2001glassy}, it is explained that
the replica symmetric assumption says that distributions $m_{v\to f}(x^{(1)},\dotsc,x^{(n)})$ and $m_{f\to v}(x^{(1)},\dotsc,x^{(n)})$ which are invariant
under permutation dominates $\mathbb{E}[Z^n]$.
Furthermore, the representations
\begin{align*}
m_{v\to f}(\pmb{x}) &= \int \prod_{i=1}^nM_{v\to f}(x_i)\mathrm{d}\Phi(M_{v\to f})\\
m_{f\to v}(\pmb{x}) &= \int \prod_{i=1}^nM_{f\to v}(x_i)\mathrm{d}\hat{\Phi}(M_{f\to v})
\end{align*}
are assumed
where $\Phi$ and $\hat{\Phi}$ denote probability measures on $\mathcal{P}(\mathcal{X})$, i.e.,
$\Phi$ and $\hat{\Phi}$ are elements of $\mathcal{P}(\mathcal{P}(\mathcal{X}))$.
Here, $\mathcal{P}(\mathcal{A})$ denotes the set of probability measures on a set $\mathcal{A}$.
\begin{lemma}\label{lem:RS}
\begin{equation*}
-F_\text{RS} = \max_{(\Phi,\hat{\Phi})\in\mathcal{S}}\left\{\frac{l}{r}\langle\log \mathcal{Z}_f\rangle
 + \langle\log \mathcal{Z}_v\rangle - l\langle\log\mathcal{Z}_{fv}\rangle\right\}
\end{equation*}
where $\mathcal{S}$ denotes the set of saddle points of the function in $\max$, where
\begin{align*}
\mathcal{Z}_v &:= \sum_{x\in\mathcal{X}}\prod_{i=1}^l M_{f\to v}^{(i)}(x)\\
\mathcal{Z}_{f} &:=  \sum_{\pmb{x}\in\mathcal{X}^r}f(\pmb{x})\prod_{i=1}^rM_{v\to f}^{(i)}(x_i)\\
\mathcal{Z}_{fv} &:= \sum_{x\in\mathcal{X}}M_{v\to f}(x)M_{f\to v}(x)
\end{align*}
where $\{M^{(i)}_{v\to f}\}_{i=1,\cdots,r}$ and $\{M^{(i)}_{f\to v}\}_{i=1,\cdots,l}$ are i.i.d. random measures obeying $\Phi$ and $\hat{\Phi}$, respectively,
and where $\langle\cdot\rangle$ denotes the expectation with respect to the random measures.
The saddle point conditions are
\begin{align*}
\frac{\prod_{i=1}^{l-1}M_{f\to v}^{(i)}(x)}{\sum_{x\in\mathcal{X}}\prod_{i=1}^{l-1}M_{f\to v}^{(i)}(x)} &\sim \Phi\\
\frac{\sum_{\pmb{x}\in\mathcal{X}^r, x_D=x}f(\pmb{x})\prod_{j=1,j\ne D}^{r}M_{v\to f}^{(j)}(x_j)}
{\sum_{\pmb{x}\in\mathcal{X}^r}f(\pmb{x})\prod_{j=1,j\ne D}^{r}M_{v\to f}^{(j)}(x_j)}
 &\sim \hat{\Phi}
\end{align*}
where $D$ denotes the uniform random variable on $\{1,2,\dotsc,r\}$ which is independent of any random variable,
and where $M\sim\Phi$ denotes that a random measure $M$ has a law $\Phi$.
\end{lemma}
Proof is in Appendix~\ref{apdx:replica}.
This derivation of replica symmetric solution is simpler than previously known ones~\cite{murayama2000statistical}, \cite{montanari2001glassy}, \cite{condamin2002study}
in which complicated tools are used e.g., integral expression of delta function.
Another advantage of this paper is that we can understand why the saddle point equation in the replica symmetric solution is equal to the DE equation.

When $f(\pmb{x})$ is invariant under permutation of $\pmb{x}$,
the fixed points for annealed free energy in Lemma~\ref{lem:stat} are also fixed point for RS saddle point equation
as delta distribution.
From inclusion relation of domains of max in Theorem~\ref{thm:ann} and Lemma~\ref{lem:RS}, $-F_\text{RS}\ge\lim_{N\to\infty}1/N\log\mathbb{E}[Z]$.
On the other hand, from Jensen's inequality, $\mathbb{E}[\log Z]\le\log\mathbb{E}[Z]$.
We now obtain the following theorem.
\begin{theorem}\label{thm:quenchann}
Assume $f(\pmb{x})$ is invariant under permutation of $\pmb{x}$.
If replica symmetric assumption is valid i.e., $-F_\text{RS}=\lim_{N\to\infty}1/N\mathbb{E}[\log Z]$,
then $\lim_{N\to\infty}1/N\mathbb{E}[\log Z] = \lim_{N\to\infty}1/N\log\mathbb{E}[Z]$.
\end{theorem}
This result is well known for regular LDPC codes~\cite{condamin2002study}.
When we believe the replica method, even if replica symmetric assumption is not valid, intuitively $-F_\text{RS} \le \lim_{N\to\infty}1/N\mathbb{E}[\log Z]$ holds,
since the replica symmetric assumption restrict the domain of maximization problem.
However, generally $-F_\text{RS}\ge \lim_{N\to\infty}1/N\mathbb{E}[\log Z]$ can be hold~\cite{guerra2003broken}.
Hence, Theorem~\ref{thm:quenchann} requires the replica symmetric assumption.

This result can be generalized for random factor model straightforwardly.
For the random magnetic field model in Appendix~\ref{apdx:mag}, $\lim_{N\to\infty}1/N\log\mathbb{E}_{\{h_i\}}[\mathbb{E}[Z^n]]$
have to be evaluated for the replica method.  
This quantity can be calculated easily by Theorem~\ref{thm:ann} by replacing $h(x)$ by $\mathbb{E}_{h}[h(x)]$. 
In this case, the relation $-F_\text{RS}\ge \lim_{N\to\infty}1/N\log\mathbb{E}_{\{h_i\}}[\mathbb{E}[Z]]$ does not hold.

\section{Applications}\label{sec:app}
\begin{figure}[t]
\psfrag{w}{$\nu(1)$}
\psfrag{G}{$\lim_{N\to\infty}\frac1N\log\mathbb{E}[Z]$}
\includegraphics[width=\hsize]{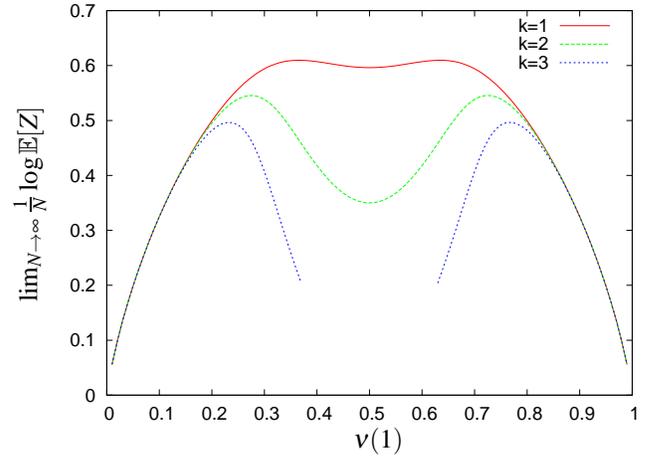}
\caption{Growth rate of the (10,20) ensembles.}
\label{fig:bcsp}
\end{figure}

In this section, an example of binary CSP is shown.
The factor is
\begin{equation}
f(\pmb{x}) = \begin{cases}
0,& \text{if } \frac{r}{2}-k < \sum_{i=1}^r x_i < \frac{r}{2}+k\\
1,& \text{otherwise}
\end{cases}
\text{for } \pmb{x}\in\{0,1\}^r.
\label{eq:bcsp}
\end{equation}
This factor is considered to prevent assignment from including half numbers of 0s and 1s.
The number of solution of fixed variable type is calculated by the BP-like algorithm shown in Section~\ref{sec:fvt}.
The calculation results for $(10,20)$ regular ensemble are shown in Fig.~\ref{fig:bcsp}.
The horizontal axis shows the relative number of 1s in solutions.
For all $k$, $\nu(1)=1/2$ is not peak of growth rate.
This means that the paramagnetic solution is not solution of the maximization problem in Theorem~\ref{thm:ann}.
For $k=3$, algorithm does not converges in region including $\nu(1)=1/2$.
When $\nu(1)=1/2$, the paramagnetic solution $m_{v\to f}(x)=m_{f\to v}(x)=1/2$ for $x=0,1$ is a fixed point of the iteration.
In Appendix~\ref{apdx:stablepara}, the stability condition of the paramagnetic solution when $\nu(1)=1/2$ is shown.
It is confirmed that the stability condition is violated for $r=20$ and $k=3$.

\section{Conclusion}
The annealed free energy of any regular, irregular and Poisson factor graph ensembles are shown.
The expression of annealed free energy includes the BP equation.
This result gives simple derivation of replica symmetric solution.
As consequence, on the replica symmetric ansatz, it is shown that annealed and quenched free energy are equal for any regular ensembles
satisfying that $f(\pmb{x})$ is invariant under permutation of $\pmb{x}$.

\section*{Acknowledgment}
The author acknowledges Toshiyuki Tanaka for insightful discussion.
This work was supported by the Grant-in-Aid for Scientific Research for JSPS Fellows (22$\cdot$5936), MEXT, Japan.

\bibliographystyle{IEEEtran}
\bibliography{IEEEabrv,ldpc}

\appendices
\section{Proof of Lemma~\ref{lem:stat}}\label{apdx:stat}
Partial derivatives of the Lagrangian~\eqref{eq:Lag} are
\begin{align*}
\frac{\partial L}{\partial \nu(x)} &= (l-1)(\log \nu(x)+1) + \lambda - l\tau(x)\\
\frac{\partial L}{\partial \mu(\pmb{x})} &= -\frac{l}{r}(\log\mu(\pmb{x})+1)+\frac{l}{r}\log f(\pmb{x}) + \frac{l}{r}\rho + \frac{l}{r}\sum_{i=1}^r\tau(x_i)
\end{align*}
We can define $m_{v\to f}(x)$ and $m_{f\to v}(x)$ which satisfies
$\sum_{x\in\mathcal{X}}m_{v\to f}(x)=
\sum_{x\in\mathcal{X}}m_{f\to v}(x)=1$ as
\begin{align*}
\tau(x) &=: \log m_{v\to f}(x) =: \log \frac1{Z_{v\to f}}m_{f\to v}(x)^{l-1}
\end{align*}
where $Z_{v\to f}$ is normalization constant.
\begin{align*}
\nu(x) &= \exp\left\{-1 - \frac{\lambda}{l-1}\right\}\left(\frac1{Z_{v\to f}}\right)^{\frac{l}{l-1}} m_{f\to v}(x)^l\\
\mu(\pmb{x}) &= \exp\left\{-1 + \rho\right\}f(\pmb{x})\prod_{i=1}^r m_{v\to f}(x)
\end{align*}
From the normalization conditions, $\lambda$ and $\rho$ are determined uniquely.
From the consistency condition of $\nu(x)$ and $\mu(\pmb{x})$, it holds
\begin{align*}
&\quad\frac1r\sum_{i=1}^r\sum_{\substack{\pmb{x}\setminus x_i\\x_i=x}} \frac1{Z_f}f(\pmb{x})\prod_{j=1}^r m_{v\to f}(x_j) = \frac1{Z_v}m_{f\to v}(x)^l\\
\Longleftrightarrow&\quad m_{v\to f}(x)\frac1r\sum_{i=1}^r\sum_{\substack{\pmb{x}\setminus x_i\\x_i=x}} \frac1{Z_f}f(\pmb{x})\prod_{j=1,j\ne i}^r m_{v\to f}(x_j)
 = \frac1{Z_v}m_{f\to v}(x)^l\\
\Longleftrightarrow&\quad \frac{Z_v}{Z_fZ_{v\to f}}\frac1r\sum_{i=1}^r\sum_{\substack{\pmb{x}\setminus x_i\\x_i=x}} f(\pmb{x})\prod_{j=1,j\ne i}^r m_{v\to f}(x_j)
 = m_{f\to v}(x).
\end{align*}

\section{Proof of Theorem~\ref{thm:ann}}\label{apdx:ann}
Let us consider
\begin{equation}\label{eq:Bethe_stat}
\frac{l}{r}\mathcal{H}(\mu)-(l-1)\mathcal{H}(v)+\frac{l}{r}\sum_{\pmb{x}\in\mathcal{X}^r}\mu({\pmb{x}})\log f(\pmb{x})
\end{equation}
evaluated at $\nu$ and $\mu$ at the stationary point i.e., they satisfy Lemma~\ref{lem:stat}.
\begin{align*}
\frac{l}{r}\mathcal{H}(\mu)&=\frac{l}{r}\log Z_f
 -\frac{l}{r}\sum_{\pmb{x}} \mu(\pmb{x})\log \left(f(\pmb{x})\prod_{i=1}^rm_{v\to f}(x_i)\right)\\
&=\frac{l}{r}\log Z_f -\frac{l}{r}\sum_{\pmb{x}}\mu(\pmb{x})\log f(\pmb{x}) - l\sum_x \nu(x)\log m_{v\to f}(x)\\
&=\frac{l}{r}\log Z_f -\frac{l}{r}\sum_{\pmb{x}}\mu(\pmb{x})\log f(\pmb{x}) - l\sum_x \nu(x)\log m_{f\to v}(x)^{l-1}\\
&\quad+l\log Z_{v\to f}
\end{align*}
\begin{align*}
(l-1)\mathcal{H}(\nu)&=(l-1)\log Z_v - (l-1)\sum_{x} \nu(x)\log m_{f\to v}(x)^l
\end{align*}
Hence,~\eqref{eq:Bethe_stat} is
\begin{equation*}
\frac{l}{r}\log Z_f + \log Z_v - l\log\frac{Z_v}{Z_{v\to f}}
\end{equation*}
The equation in Theorem~\ref{thm:ann} is obtained from the equality
\begin{equation*}
\frac{Z_v}{Z_{v\to f}}
 = \frac{\sum_{x\in\mathcal{X}}m_{f\to v}(x)^l}{\sum_{z\in\mathcal{X}}m_{f\to v}(z)^{l-1}}
 = Z_{fv}.
\end{equation*}

On the other hand, let us consider the function
\begin{equation*}
F(\{m_{v\to f}\},\{m_{f\to v}\}) := \frac{l}{r}\log Z_f + \log Z_v - l\log Z_{fv}
\end{equation*}
for any non-negative functions $m_{f\to v}(x)$ and $m_{v\to f}(x)$.
This quantity is invariant under scaling of $\{m_{f\to v}(x)\}_{x\in\mathcal{X}}$ and $\{m_{v\to f}(x)\}_{x\in\mathcal{X}}$.
Hence, without loss of generality, we can assume $\sum_x m_{f\to v}(x)=\sum_x m_{v\to f}(x)=1$.
Since the first derivatives are
\begin{align*}
\frac{\partial F(\{m_{v\to f}\},\{m_{f\to v}\})}{\partial m_{f\to v}(x)} &= l\frac{m_{f\to v}(x)^{l-1}}{Z_v}-l\frac{m_{v\to f}(x)}{Z_{fv}}\\
\frac{\partial F(\{m_{v\to f}\},\{m_{f\to v}\})}{\partial m_{v\to f}(x)} &= \frac{l}{r}\frac{
\sum_{i=1}^r\sum_{\substack{\pmb{x}\setminus x_i\\x_i=x}}f(\pmb{x})\prod_{j\ne i}m_{v\to f}(x_j)
}{Z_f}\\
&\qquad-l\frac{m_{f\to v}(x)}{Z_{fv}}
\end{align*}
the saddle point condition is
\begin{align*}
m_{v\to f}(x)&=\frac{Z_{fv}}{Z_v} m_{f\to v}(x)^{l-1}\\
m_{f\to v}(x)&=\frac1r\frac{Z_{fv}}{Z_f} \sum_{i=1}^r\sum_{\substack{\pmb{x}\setminus x_i\\x_i=x}}f(\pmb{x})\prod_{j=1, j\ne i}^r m_{v\to f}(x_j).
\end{align*}
Although these points take minimal on any lines parallel to axis of $\{m_{f\to v}(x)\}$ or $\{m_{v\to f}(x)\}$,
these points are not necessarily minimal.

\section{Magnetic field model}\label{apdx:mag}
Although we have only considered the random regular factor graph ensembles,
the method can be generalized straightforwardly to many ensembles.
As a simple example, we introduce the random regular factor graph with magnetic field.
\begin{equation*}
p(\pmb{x}) = \frac1Z\prod_a f(\pmb{x}_{\partial a}) \prod_i h(x_i).
\end{equation*}
Here, there are degree one factor nodes for each variable node.
In the same way, the annealed free energy can be calculated.
\begin{multline*}
\lim_{N\to\infty} \frac1N \log \mathbb{E}[Z(\{\nu\},\{\mu\})]\\
= \frac{l}{r}\mathcal{H}(\{\mu\}) - (l-1)\mathcal{H}(\{\nu\})
+\frac{l}{r}\sum_{\pmb{x}\in\mathcal{X}^r} \mu(\pmb{x})\log f(\pmb{x})\\
+\sum_x\nu(x)\log h(x).
\end{multline*}
\begin{lemma}\label{lem:mag}
\begin{multline*}
\lim_{N\to\infty}\frac1N\log\mathbb{E}[Z]\\
 = \max_{(m_{f\to v}(x),m_{v\to f}(x))\in\mathcal{S}}\left\{\frac{l}{r}\log Z_f+\log Z_v - l\log Z_{fv}\right\}
\end{multline*}
where $\mathcal{S}$ denotes the set of saddle points of the function in $\max$, and where
\begin{align*}
Z_{v} &:= \sum_{x} h(x)m_{f\to v}(x)^l\\
Z_{f} &:= \sum_{\pmb{x}} f(\pmb{x}) \prod_{i=1}^r m_{v\to f}(x_i)\\
Z_{fv} &:= \sum_{x} m_{f\to v}(x)m_{v\to f}(x)
\end{align*}
The stationary condition is
\begin{align*}
m_{v\to f}(x)&\propto h(x)m_{f\to v}(x)^{l-1}\\
m_{f\to v}(x)&\propto \sum_{i=1}^r\sum_{\substack{\pmb{x}\setminus x_i\\x_i=x}}f(\pmb{x})\prod_{j=1, j\ne i}^r m_{v\to f}(x_j).
\end{align*}
\end{lemma}
The above stationary condition is related to the stationary condition of maximization of the contribution $Z(\nu,\mu)$ with fixed variable type $\nu$ in Section~\ref{sec:fvt}. 

\section{Proof of Lemma~\ref{lem:vtypestat}}\label{apdx:vtypestat}
Generally, when we have additional linear constraints
\begin{align*}
\sum_{x\in\mathcal{X}}a_k(x)\nu(x) &= b_k, \hspace{2em} \text{for } k = 1,2,\dotsc,s\\
\sum_{\pmb{x}^r\in\mathcal{X}}c_k(\pmb{x})\mu(\pmb{x}) &= d_k, \hspace{2em} \text{for } k = 1,2,\dotsc,t
\end{align*}
in the maximization problem of $-F_\text{Bethe}(\{\nu\},\{\mu\})$,
the stationary condition is
\begin{align*}
\mu(\pmb{x}) &\propto \left(\prod_{k=1}^tg_k^{c_k(\pmb{x})}\right)f(\pmb{x})\prod_{i=1}^r m_{v\to f}(x_i)\\
\nu(x) &\propto \left(\prod_{k=1}^sh_k^{a_k(x)}\right)m_{f\to v}(x)^l\\
m_{v\to f}(x)&\propto \left(\prod_{k=1}^sh_k^{a_k(x)}\right)m_{f\to v}(x)^{l-1}\\
m_{f\to v}(x)&\propto \sum_{i=1}^r\sum_{\substack{\pmb{x}\setminus x_i\\x_i=x}}\left(\prod_{k=1}^tg_k^{c_k(\pmb{x})}\right)f(\pmb{x})\prod_{j=1, j\ne i}^r m_{v\to f}(x_j).
\end{align*}
where $\{h_k\ge 0\}_{k=1,\dotsc,s}$ and $\{g_k\ge 0\}_{k=1,\dotsc,t}$ are auxiliary variables.
\begin{IEEEproof}
\begin{multline*}
L(\nu,\mu;\lambda,\rho, \eta, \zeta, \tau) = -F_\text{Bethe}\\
+ \lambda\left(\sum_{x\in\mathcal{X}}\nu(x)-1\right)
+ \sum_{k=1}^s \eta_k\left(\sum_{x\in\mathcal{X}}a_{k}(x)\nu(x) - b_k\right)\\
+ \frac{l}{r}\rho\left(\sum_{\pmb{x}\in\mathcal{X}^r}\mu(\pmb{x})-1\right)
+ \frac{l}{r}\sum_{k=1}^t \zeta_k\left(\sum_{\pmb{x}\in\mathcal{X}^r}c_k(\pmb{x})\mu(\pmb{x}) - d_k\right)\\
+\sum_{z\in\mathcal{X}}\tau(z)\left(\frac{l}{r}\sum_{i=1}^r\sum_{\substack{\pmb{x}\setminus x_i\\ x_i=z}}\mu(\pmb{x}) - l\nu(z)\right).
\end{multline*}
\begin{align*}
\frac{\partial L}{\partial \nu(x)} &= (l-1)(\log \nu(x)+1) + \lambda + \sum_{k=1}^s\eta_ka_k(x) - l\tau(x)\\
\frac{\partial L}{\partial \mu(\pmb{x})} &= -\frac{l}{r}(\log\mu(\pmb{x})+1)+\frac{l}{r}\log f(\pmb{x}) + \frac{l}{r}\rho\\
&\quad + \frac{l}{r}\sum_{k=1}^t\zeta_kc_k(\pmb{x}) + \frac{l}{r}\sum_{i=1}^r\tau(x_i).
\end{align*}
Let 
\begin{align*}
\tau(x) &=: \log m_{v\to f}(x) =: \log\left( \frac1{Z_{v\to f}}\left(\prod_{k=1}^sh_k^{a_k(x)}\right)m_{f\to v}(x)^{l-1}\right)\\
\eta_k &=: \log h_k\\
\zeta_k &=: \log g_k.
\end{align*}
The rest of the proof is same as the proof of Lemma~\ref{lem:stat}.
\end{IEEEproof}

\section{Proof of Lemma~\ref{lem:RS}}\label{apdx:replica}
We use the following relation.
\begin{equation*}
\lim_{n\to 0}\frac1n \log\langle A^n\rangle
=\langle \log A \rangle
\end{equation*}
where $A$ is a random variable and $\langle\cdot\rangle$ denotes an expectation.
\begin{align*}
Z_v &= \sum_{\pmb{x}\in\mathcal{X}^n}\left(\int\prod_{i=1}^n M_{v\to f}(x_i)\mathrm{d}P\right)^l\\
&= \int\dotsm\int \left(\prod_{j=1}^l\mathrm{d}P_j\right)\left(\sum_{x}\prod_{j=1}^l M^{(j)}_{v\to f}(x)\right)^n
\end{align*}
Hence,
\begin{equation*}
\lim_{n\to 0}\frac1n \log Z_v = \langle \log\mathcal{Z}_v\rangle
\end{equation*}
The derivation of $\mathcal{Z}_f$ and $\mathcal{Z}_{fv}$ are similar.

The derivation of the saddle point equations are omitted since it is straightforward.

\section{Regular LDPC codes}
\begin{corollary}
[(Litsyn and Shevelev, 2002), (Burshtein and Miller, 2004)]
Growth rate of $(l,r)$-regular LDPC code ensemble is
\begin{multline*}
G(\omega) = 
\frac{l}{r}\log\frac{1+{z'}^r}2\\
+ \log\left[\mathrm{e}^h\left(\frac{1+y'}2\right)^l + \mathrm{e}^{-h}\left(\frac{1-y'}2\right)^l\right]\\
-l\log\frac{1+y'z'}2 - \omega'h
\end{multline*}
where $\omega':=1-2\omega$ and
\begin{align*}
\omega' &= \tanh(h+l\tanh^{-1}(y'))\\
y' &= z'^{r-1}\\
z' &= \tanh(h+(l-1)\tanh^{-1}(y')).
\end{align*}
\end{corollary}
This result can be easily understood from Lemma~\ref{lem:vtypestat} and~\ref{lem:vtype} by observing the following correspondings,
\begin{align*}
\omega' &= \nu(0)-\nu(1),&
h &= (-1)^{x}\log h(x)\\
z' &= m_{v\to f}(0) - m_{v\to f}(1),&
y' &= m_{f\to v}(0) - m_{f\to v}(1)
\end{align*}
and
\begin{align*}
Z_f &= \log \frac{1+{z'}^r}2\\
Z_v &= \log\left[\mathrm{e}^h\left(\frac{1+y'}2\right)^l + \mathrm{e}^{-h}\left(\frac{1-y'}2\right)^l\right]\\
Z_{fv} &= \log\frac{1+y'z'}2\\
\sum_x\nu(x)\log h(x) &= \omega'h.
\end{align*}
This result is also obtained by using the combinatorial method in~\cite{mezard2009information} and change of variables~\cite{di2005weight}
\begin{align*}
h &= -\frac12 \log x,&
y'&= \frac{1-y}{1+y},&
z'&= \frac{1-z}{1+z}.
\end{align*}
But the proof of this paper is much more meaningful.

\section{Random magnetic field model}
In this appendix, we consider the random magnetic field model.
\begin{align*}
p(\pmb{x}\mid \{h_i\}) &= \frac1{Z(\{h_i\})}\prod_a f(\pmb{x}_{\partial a}) \prod_i h_i(x_i)\\
Z(\{h_i\}) &= \sum_{\pmb{x}}\prod_a f(\pmb{x}_{\partial a}) \prod_i h_i(x_i).
\end{align*}
Here, $\{h_i\}$ independently and identically distributed according to the distribution $P_H(h)$ on a finite set $\mathcal{H}$ of nonnegative function on $\mathcal{X}$.
In statistical physics, $h_i(x_i)$ represents random magnetic field.
As a posteriori probability of LDPC codes, $h_i$ corresponds to output of a channel.
We now consider $\lim_{N\to\infty}1/N\mathbb{E}_{\{h_i\}}[\log\mathbb{E}[Z(\{h_i\})]]$.
Since $\mathbb{E}[Z(\{h_i\})]$ depends on $\{h_i\}$ only through the type of $\{h_i\}$, and since $1/N\log\mathbb{E}[Z(\{h_i\})]=O(1)$
for any $\{h_i\}$, we only have to deal with typical $\{h_i\}$.
Let $v(x,h)$ denotes the number of variable nodes of value $x$ and whose corresponding factor is $h$.
The factor-type $u(\pmb{x},\pmb{h})$ is defined in the same way.
Then, it holds
\begin{align*}
Z &= \sum_{v,u} N(v,u)\prod_{(\pmb{x},\pmb{h})\in\mathcal{X}^r\times\mathcal{H}^r}f(\pmb{x})^{u(\pmb{x},\pmb{h})}\prod_{(x,h)\in\mathcal{X}\times\mathcal{H}} h(x)^{v(x,h)}
\end{align*}
For typical $\{h_i\}$, it holds
\begin{multline*}
\mathbb{E}[N(v,u)] = \prod_{h\in\mathcal{H}}\binom{NP_H(h)}{\{v(x,h)\}_{x\in\mathcal{X}}}
\binom{\frac{l}{r}N}{\{u(\pmb{x},\pmb{h})\}_{(\pmb{x},\pmb{h})\in\mathcal{X}^r\times\mathcal{H}^r}}\\
\cdot\frac{\prod_{(x,h)\in\mathcal{X}\times\mathcal{H}}(v(x,h)l)!}{(Nl)!}.
\end{multline*}

Hence, the problem is maximization of
\begin{multline*}
\frac{l}{r}\mathcal{H}(\mu)-(l-1)\mathcal{H}(\nu) - \mathcal{H}(P_H)\\
+\frac{l}{r}\sum_{(\pmb{x},\pmb{h})\in\mathcal{X}^r\times\mathcal{H}^r}\mu(\pmb{x},\pmb{h})\log f(\pmb{x})
+\sum_{(x,h)\in\mathcal{X}\times\mathcal{H}}\nu(x,h)\log h(x)
\end{multline*}
subject to
\begin{align*}
\nu(x,h)&\ge 0,&
\mu(\pmb{x},\pmb{h})&\ge 0\\
\sum_{x}\nu(x,h)&= P_H(h),&
\sum_{\pmb{x},\pmb{h}}\mu(\pmb{x},\pmb{h})&= 1
\end{align*}
\begin{equation*}
\frac1{r}\sum_{i=1}^r\sum_{\substack{(\pmb{x},\pmb{h})\setminus (x_i,h_i)\\ x_i=z,h_i=h}}\mu(\pmb{x},\pmb{h}) = \nu(z,h),\forall z\in\mathcal{X}, h\in\mathcal{H}.
\end{equation*}

\begin{lemma}
The stationary conditions are
\begin{align*}
\mu(\pmb{x},\pmb{h})&\propto f(\pmb{x})\prod_{i=1}^r m_{v\to f}(x_i, h_i)\\
\nu(x,h)&\propto g(h)h(x) m_{f\to v}(x)^l
\end{align*}
where
\begin{align}
P_H(h) &\propto g(h)\sum_{x\in\mathcal{X}} h(x)m_{f\to v}(x)^l\label{eq:rfmstats}\\
m_{v\to f}(x,h)&\propto g(h)h(x)m_{f\to v}(x)^{l-1}\nonumber\\
m_{v\to f}(x)&\propto \sum_h m_{v\to f}(x,h)\\
m_{f\to v}(x)&\propto \sum_{i=1}^r\sum_{\substack{\pmb{x}\setminus x_i\\x_i=x}}f(\pmb{x})\prod_{j=1, j\ne i}^r m_{v\to f}(x_j).\label{eq:rfmstate}
\end{align}
\end{lemma}
Here $m_{v\to f}(x)$, $m_{f\to v}(x)$ and $g(h)$ are auxiliary functions satisfying
$\sum_{x\in\mathcal{X}} m_{v\to f}(x) = \sum_{x\in\mathcal{X}} m_{f\to v}(x) = 1$,
and $g(h)\ge 0$.

\begin{lemma}
\begin{align*}
&\lim_{N\to\infty}\mathbb{E}_{\{h_i\}}[\log\mathbb{E}[Z]]\\
&\quad = \max_{(m_{f\to v}(x),m_{v\to f}(x),g(h))\in\mathcal{S}}
\Bigg\{\frac{l}{r}\log Z_f + \log Z_v - l\log Z_{fv}\\
&\qquad + \sum_{h}P_H(h)\log \frac{P_H(h)}{g(h)}\Bigg\}\\
&\quad = \max_{(m_{f\to v}(x),m_{v\to f}(x))\in\mathcal{S}}\Bigg\{\frac{l}{r}\log Z_f + \sum_{h} P_H(h)\log Z_v(h)\\
&\qquad - l\log Z_{fv}\Bigg\}
\end{align*}
where $\mathcal{S}$ denotes the set of saddle points of the function in $\max$, and where
\begin{align*}
Z_f &:= \sum_{\pmb{x}\in\mathcal{X}^r}f(\pmb{x})\prod_{i=1}^r m_{v\to f}(x_i)\\
Z_v(h) &:= \sum_{x\in\mathcal{X}} h(x)m_{f\to v}(x)^l\\
Z_v &:= \sum_{h\in\mathcal{H}}g(h)Z_v(h)\\
Z_{fv} &:= \sum_{x\in\mathcal{X}} m_{f\to v}(x)m_{v\to f}(x).
\end{align*}
The conditions of saddle point are~\eqref{eq:rfmstats} to \eqref{eq:rfmstate}.
\end{lemma}

\section{Irregular and Poisson ensembles}\label{apdx:irgpoi}
\subsection{Irregular ensemble}
The result can be generalized for irregular ensembles.
Let $\mathcal{D}_v$ and $\mathcal{D}_c$ denote the set of degrees of variable nodes and check nodes, respectively.
Let $L_i$ and $R_j$ denote the degree distribution of variable nodes and check nodes from node perspective for $i\in\mathcal{D}_v$ and $j\in\mathcal{D}_c$, respectively.
Assume the factor corresponding to degree $j$ factor nodes is $f_j(\pmb{x})$ for $j\in\mathcal{D}_c$.
Let $v(i,x)$ denotes the number of variable nodes of degree $i$ and value $x$.
The factor-type $u(j,\pmb{i},\pmb{x})$ is defined in the same way.
\begin{multline*}
\mathbb{E}[N(v,u)] =
\prod_{i\in\mathcal{D}_v} \binom{NL_i}{\{v(i,x)\}_{x\in\mathcal{X}}}\\
\times\prod_{j\in\mathcal{D}_c}\binom{\frac{L'(1)}{R'(1)}NR_j}{\{u_{j,\pmb{i},\pmb{x}}\}_{(\pmb{i},\pmb{x})\in\mathcal{D}_v^j\times\mathcal{X}^j}}
\frac{\prod_{(i,x)\in\mathcal{D}_v\times\mathcal{X}}(v(i,x)i)!}{(NL'(1))!}
\end{multline*}
The problem is the maximization of
\begin{align*}
&\lim_{N\to\infty} \frac1N \log \mathbb{E}[Z(\nu,\mu)]\\
&= \frac{L'(1)}{R'(1)}\sum_{j\in\mathcal{D}_c}R_j\mathcal{H}(\mu_j) - 
\sum_{i\in\mathcal{D}_v}L_i(i-1)\mathcal{H}(\nu_i) - L'(1)\mathcal{H}(L_ii)\\
&\quad+\frac{L'(1)}{R'(1)}\sum_{j\in\mathcal{D}_c}\sum_{(\pmb{i},\pmb{x})\in\mathcal{D}_v^j\times\mathcal{X}^j} \mu(j,\pmb{i},\pmb{x})\log f_j(\pmb{x})
\end{align*}
subject to
\begin{align*}
\nu(i,x) &\ge 0,& \mu(j,\pmb{i},\pmb{x})&\ge 0\\
\sum_{x\in\mathcal{X}}\nu(i,x) &= L_i,&
\sum_{(\pmb{i},\pmb{x})\in\mathcal{D}_v^j\times\mathcal{X}^j}\mu(j,\pmb{i},\pmb{x}) &= R_j
\end{align*}
\begin{equation*}
\frac{L'(1)}{R'(1)}\sum_{j\in\mathcal{D}_c}\sum_{k=1}^j\sum_{(\pmb{i},\pmb{x}), (i_k, x_k) = (i,x)}\mu(j,\pmb{i},\pmb{x}) = i\nu(i,x).
\end{equation*}
We obtain the following stationary conditions.
\begin{align}
\mu(j,\pmb{i},\pmb{x}) &\propto r(j)f_j(\pmb{x}) \prod_{k=1}^j m_{v\to f}(i_k,x_k)\nonumber\\
\nu(i,x) &\propto l(i)m_{f\to v}(x)^{i}\nonumber\\
L_i &\propto l(i)\sum_{x\in\mathcal{X}}m_{f\to v}(x)^i\label{eq:irgstats}\\
R_j &\propto r(j)\sum_{\pmb{x}\in\mathcal{X}^j}f_j(\pmb{x})\prod_{k=1}^jm_{v\to f}(x_k)\\
m_{v\to f}(i,x)&\propto il(i)m_{f\to v}(x)^{i-1}\nonumber\\
m_{v\to f}(x)&\propto \sum_{i\in\mathcal{D}_v}il(i)m_{f\to v}(x)^{i-1}\\
m_{f\to v}(x)&\propto \sum_{j\in\mathcal{D}_c}\sum_{t=1}^j\sum_{\substack{\pmb{x}\in\mathcal{X}^j,\pmb{x}\setminus x_t\\x_t=x}}r(j)f_j(\pmb{x})
\prod_{k=1, k\ne t}^j m_{v\to f}(x_k)\label{eq:irgstate}
\end{align}

\begin{lemma}
\begin{align}
&\lim_{N\to\infty}\frac1N\log\mathbb{E}[Z]\nonumber\\
&\quad=\max_{(m_{v\to f}(x),m_{f\to v}(x),l(i),r(j))\in\mathcal{S}}\Bigg\{\frac{L'(1)}{R'(1)} \log Z_f 
+\log Z_v\nonumber\\
&\quad\quad - L'(1)\log Z_{fv}+\frac{L'(1)}{R'(1)} \sum_{j\in\mathcal{D}_c} R_j\log \frac{R_j}{r(j)} + \sum_{i\in\mathcal{D}_v}L_i\log \frac{L_i}{l(i)}\Bigg\}\nonumber\\
&\quad=\max_{(m_{v\to f}(x),m_{f\to v}(x))\in\mathcal{S}}\Bigg\{\frac{L'(1)}{R'(1)}\sum_{j\in\mathcal{D}_c}\log Z_f(j) + \sum_{i\in\mathcal{D}_v} L_i \log Z_v(i)\nonumber\\
&\quad\quad - L'(1)\log Z_{fv}\Bigg\}\label{eq:irg}
\end{align}
where $\mathcal{S}$ denotes the set of saddle points of the function in $\max$, and where
\begin{align*}
Z_f(j) &:= \sum_{\pmb{x}\in\mathcal{X}^j}f_j(\pmb{x})\prod_{k=1}^jm_{v\to f}(x_k)\\
Z_v(i) &:= \sum_{x\in\mathcal{X}}m_{f\to v}(x)^i\\
Z_v &:= \sum_{i\in\mathcal{D}_v} l(i) Z_v(i)\\
Z_f &:= \sum_{j\in\mathcal{D}_c} r(j)Z_f(j)\\
Z_{fv} &:= \sum_{x\in\mathcal{X}}m_{f\to v}(x)m_{v\to f}(x)
\end{align*}
The stationary conditions are \eqref{eq:irgstats} to \eqref{eq:irgstate}. 
\end{lemma}
This second expression~\eqref{eq:irg} is equivalent to the equations in~\cite{di2005weight}.

\subsection{Poisson ensemble}
In this subsection, we deal with Poisson ensemble.
There are $N$ variable nodes and $\alpha N$ factor nodes.
The degree of factor node is $k$.
For each factor node, connecting variable nodes are chosen independently and uniformly from $N(N-1)\dotsm (N-(k-1))$ ways.
In the same way as other ensembles, we obtain
\begin{multline*}
\mathbb{E}[N(v,u)]
=
\binom{N}{\{v(x)\}_{x\in\mathcal{X}}}
\binom{\alpha N}{\{u(\pmb{x})\}_{\pmb{x}\in\mathcal{X}^k}}\\
\times\prod_{\pmb{x}\in\mathcal{X}^k}\left(\frac{\prod_{x\in\mathcal{X}}v(x)(v(x)-1)\dotsm(v(x)-(N_x(\pmb{x})-1))}{N(N-1)\dotsm(N-(k-1))}\right)^{u(\pmb{x})}
\end{multline*}
where $N_x(\pmb{x})$ denotes the number of $x$ in $\pmb{x}$.
The problem is maximization of
\begin{align*}
&\lim_{N\to\infty} \frac1N \log \mathbb{E}[Z(\nu,\mu)]\\
&\quad= \alpha\mathcal{H}(\mu) 
+\mathcal{H}(\nu)
+\alpha\sum_{\pmb{x}\in\mathcal{X}^k}\mu(\pmb{x})\log\left(\prod_{i=1}^k\nu(x_i)\right)\\
&\quad\quad+\alpha\sum_{\pmb{x}\in\mathcal{X}^k} \mu(\pmb{x})\log f(\pmb{x})\\
&\quad=-\alpha\mathcal{D}(\mu\| \nu^k)+\mathcal{H}(\nu)+\alpha\sum_{\pmb{x}\in\mathcal{X}^k}\mu(\pmb{x})\log f(\pmb{x})
\end{align*}
subject to
\begin{align*}
\nu(x)&\ge 0,& \mu(\pmb{x})&\ge 0\\
\sum_x\nu(x)&= 1,& \sum_{\pmb{x}}\mu(\pmb{x})&=1.
\end{align*}
This is also similar to the minimization of Bethe free energy since \eqref{eq:Betheent} is also written as
\begin{align*}
\mathcal{H}_\text{Bethe}(b_i,b_a) &= -\sum_{a}\sum_{\pmb{x}_{\partial a}} b_a(\pmb{x}_{\partial a}) \log \frac{b_a(\pmb{x}_{\partial a})}{\prod_{j\in\partial a}b_j(x_j)}\\
&\quad -\sum_{i}\sum_{x_i} b_i(x_i) \log b_i(x_i).
\end{align*}
The derivation of the following lemma is omitted for lack of space.
\begin{lemma}
\begin{align}
&\lim_{N\to\infty} \frac1N \log \mathbb{E}[Z(\nu,\mu)]\nonumber\\
&=\max_{(m_{f\to v}(x),m_{v\to f}(x),e)\in\mathcal{S}}\Bigg\{\alpha\log Z_f+\log Z_v\nonumber\\
&\hspace{10em} - e\sum_{x}m_{v\to f}(x)m_{f\to v}(x)\Bigg\}\label{eq:poi}
\end{align}
where $\mathcal{S}$ denotes the set of saddle points of the function in $\max$, and where
\begin{align*}
Z_f &:= \sum_{\pmb{x}}f(\pmb{x})\prod_{i=1}^km_{v\to f}(x_i)\\
Z_v &:= \sum_{x}\exp\{em_{f\to v}(x)\}
\end{align*}
The conditions of saddle point are
\begin{align*}
m_{f\to v}(x) &=\frac{\alpha}{eZ_f} \sum_{i=1}^k\sum_{\substack{\pmb{x}\in\mathcal{X}^k\\x_i=x}} f(\pmb{x})\prod_{j=1, j\ne i}^k m_{v\to f}(x_j)\\
m_{v\to f}(x) &\propto \exp \{em_{f\to v}(x)\}\\
&= 1 + em_{f\to v}(x) + \frac{(em_{f\to v}(x))^2}{2!} + \dotsb.
\end{align*}
\end{lemma}
Here, $e$ can be regarded as mean of Poisson distribution expressing the degree distribution of variable nodes.
Note that the third term of~\eqref{eq:poi} evaluated at saddle points is $\alpha k$.

\section{Stability of the paramagnetic solution}\label{apdx:stablepara}
Assume $\sum_{i=1}^r\sum_{\substack{\pmb{x}\setminus x_i\\x_i=x}}f(\pmb{x})$ is constant among all $x\in\mathcal{X}$.
Let $\nu(x)=m_{f\to v}^P(x) = m_{v\to f}^P(x) = 1/q$ for all $x\in\mathcal{X}$.
Let us start the algorithm in Section~\ref{sec:fvt} from
\begin{equation*}
m_{f\to v}(x)\propto m_{f\to v}^P(x) + \delta(x).
\end{equation*}
By linear approximation,
\begin{align*}
m_{v\to f}^+(x) &\propto \frac{\nu(x)}{m_{f\to v}^P(x)+\delta(x)}\\
&= \frac{\nu(x)}{m_{f\to v}^P(x)}\left[1-\frac{\delta(x)}{m_{f\to v}^P(x)}+\Theta(\delta(x)^2)\right]\\
&= 1 -q\delta(x)+\Theta(\delta(x)^2)\\
m_{f\to v}^+(x)&\propto \sum_{i=1}^r\sum_{\substack{\pmb{x}\setminus x_i\\x_i=x}}f(\pmb{x})\prod_{j=1, j\ne i}^r m_{v\to f}^+(x_j)\\
&\propto \sum_{i=1}^r\sum_{\substack{\pmb{x}\setminus x_i\\x_i=x}}f(\pmb{x})\prod_{j=1, j\ne i}^r\left(1 -q\delta(x_j)+ \Theta(\delta(x_j)^2)\right)\\
&= \sum_{i=1}^r\sum_{\substack{\pmb{x}\setminus x_i\\x_i=x}}f(\pmb{x})
- q\sum_{i=1}^r\sum_{\substack{\pmb{x}\setminus x_i\\x_i=x}}f(\pmb{x})\left(\sum_{j=1, j\ne i}^r\delta(x_j)\right) + \sum_{x\in\mathcal{X}}\Theta(\delta(x)^2)\\
&\propto \frac1q - 
\frac{\sum_{i=1}^r\sum_{\substack{\pmb{x}\setminus x_i\\x_i=x}}f(\pmb{x})\left(\sum_{j=1, j\ne i}^r\delta(x_j)\right)}
{\sum_{i=1}^r\sum_{\substack{\pmb{x}\setminus x_i\\x_i=x}}f(\pmb{x})}
 + \sum_{x\in\mathcal{X}}\Theta(\delta(x)^2)
\end{align*}
Let
\begin{equation*}
\delta^+(x) :=
-\frac{\sum_{i=1}^r\sum_{\substack{\pmb{x}\setminus x_i\\x_i=x}}f(\pmb{x})\left(\sum_{j=1, j\ne i}^r\delta(x_j)\right)}
{\sum_{i=1}^r\sum_{\substack{\pmb{x}\setminus x_i\\x_i=x}}f(\pmb{x})}.
\end{equation*}
We now consider the linear operator $A$ defined by $A(\{\delta(x)\}_{x\in\mathcal{X}})=\{\delta^+(x)\}_{x\in\mathcal{X}}$.
The all 1 vector is a eigenvector of $A$ with eigenvalue $-(r-1)$.
The stability condition of the paramagnetic solution is that absolute values of eigenvalues of $A$ not corresponding to the all 1 vector are smaller than 1.
For the binary CSP~\eqref{eq:bcsp}, the matrix $A$ is a symmetric $2\times 2$ matrix where
\begin{align*}
A_{11}=A_{22}&=-\frac{(r-1)\sum_{i=0}^{\frac{r}{2}-k-1}\binom{r-1}{i} + \binom{r-1}{\frac{r}{2}-k}\left(\frac{r}{2}+k-1\right)}
{2\sum_{i=0}^{\frac{r}{2}-k-1}\binom{r-1}{i}+\binom{r-1}{\frac{r}{2}-k}}\\
A_{12}=A_{21}&=-\frac{(r-1)\sum_{i=0}^{\frac{r}{2}-k-1}\binom{r-1}{i} + \binom{r-1}{\frac{r}{2}-k}\left(\frac{r}{2}-k\right)}
{2\sum_{i=0}^{\frac{r}{2}-k-1}\binom{r-1}{i}+\binom{r-1}{\frac{r}{2}-k}}.
\end{align*}
The eigenvalues of $A$ are $A_{11}+A_{12}$ and $A_{11}-A_{12}$ whose eigenvectors are $[1\; 1]^T$ and $[1\; -1]^T$, respectively.
We can easily confirm that
\begin{align*}
A_{11}+A_{12} &= -(r-1)\\
A_{11}-A_{12} &= 
-\frac{\binom{r-1}{\frac{r}{2}-k}(2k-1)}
{2\sum_{i=0}^{\frac{r}{2}-k-1}\binom{r-1}{i}+\binom{r-1}{\frac{r}{2}-k}}.
\end{align*}
Hence, the stability condition is
\begin{equation*}
\frac{\binom{r-1}{\frac{r}{2}-k}(2k-1)}
{2\sum_{i=0}^{\frac{r}{2}-k-1}\binom{r-1}{i}+\binom{r-1}{\frac{r}{2}-k}}
<1.
\end{equation*}
For $r=20$ and $k=1,2,3$, the left-hand side of the condition is $0.23883$, $0.859049$ and $1.825917$, respectively.
This result is consistent with the numerical calculation result in Fig.~\ref{fig:bcsp}.

\end{document}